\documentclass[twocolumn]{jpsj2} %% two-column layout
\title{Effect of Oscillating Landau Bandwidth on the Integer Quantum Hall Effect in a Unidirectional Lateral Superlattice}

\author{Akira \textsc{Endo} \thanks{E-mail: akrendo@issp.u-tokyo.ac.jp} and Yasuhiro \textsc{Iye}}

\inst{Institute for Solid State Physics, University of Tokyo, 5-1-5 Kashiwanoha, Kashiwa, Chiba 277-8581, Japan}

\abst{
We have measured activation gaps for odd-integer quantum Hall states in a unidirectional lateral superlattice (ULSL) --- a two-dimensional electron gas (2DEG) subjected to a unidirectional periodic modulation of the electrostatic potential. By comparing the activation gaps with those simultaneously measured in the adjacent section of the same 2DEG sample without modulation, we find that the gaps are reduced in the ULSL by an amount corresponding to the width acquired by the Landau levels through the introduction of the modulation. The decrement of the activation gap varies with the magnetic field following the variation of the Landau bandwidth due to the commensurability effect. Notably, the decrement vanishes at the flat band conditions.   
}

\kword{integer quantum Hall effect, unidirectional lateral superlattice, potential modulation, Landau bandwidth, activation gap, Zeeman energy}

\begin{document}
\maketitle

\section{Introduction\label{Introduction}} %% No sections necessary for express letters, letters and short notes
\begin{figure}
\includegraphics[bbllx=20,bblly=20,bburx=550,bbury=370,width=8.5cm]{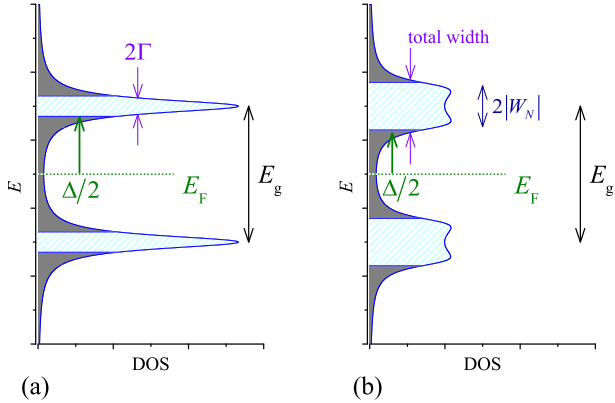}%
\caption{(Color online) Sketch of the density of states (DOS) comprised of discrete disorder-broadened Landau levels for (a) a plain 2DEG with the Landau peaks given by a Lorentzian, eq.\ (\ref{LorentzP}), and (b) a ULSL with the Landau peaks given by eq.\ (\ref{VLorP}). Dark (gray) and bright (light blue) shaded regions represent localized and extended states, respectively. Relevant energies are also noted in the figure; $E_\mathrm{F}$: the Fermi energy (for an integer filling of the Landau levels), $E_\mathrm{g}$: intrinsic gap between the levels, $\Gamma$: disorder broadening, $|W_N|$: width introduced by the modulation, $\Delta /2$: activation energy.  \label{LLs}}
\end{figure}

A crystal that hosts a two-dimensional electron gas (2DEG) inevitably contains a certain degree of disorder. It is worth noting that the disorder is an indispensable ingredient for the occurrence of the integer quantum Hall effect (IQHE) \cite{PrangeGirvin90,Janssen94,ChakrabortyPietilainen95}. Disorder introduces random fluctuations, both in magnitude and spatial distribution, in the electrostatic potential experienced by the electrons in the 2DEG\@. Owing to the random potential, Landau levels acquire width, and the tails of the Landau peaks fill the gap between adjacent levels with low-density spatially localized states. The resultant density of states (DOS) are schematically depicted in Fig.\ \ref{LLs} (a). At an integer filling of the Landau levels, the Fermi energy $E_\mathrm{F}$ lies at the localized states, resulting in the quantum Hall state characterized by a vanishing longitudinal conductivity $\sigma_{xx}$ in the low-temperature limit, as well as by a plateau in the transverse conductivity $\sigma_{xy}$.

At elevated temperatures, $\sigma_{xx}$ revives with the electrons thermally activated across the energy gap to spatially extended states lying above the mobility edge, giving rise to the temperature dependence $\sigma_{xx} \propto \exp \left(-{\Delta}/{2 k_\mathrm{B} T} \right)$ \cite{PrangeGirvin90,NoteVRH,Furlan98}. Here again, the random potential plays an important role: the activation energy $\Delta /2$ is reduced from $E_\mathrm{g}/2$, with $E_\mathrm{g}$ representing the intrinsic gap (the energy difference between adjacent levels, see Fig.\ \ref{LLs}), by an amount basically corresponding to the width $\Gamma$ acquired by the Landau level. Consequently, IQH states are made less robust by the random potential. It is interesting to point out that the random potential plays two seemingly conflicting roles of giving birth to and disturbing the IQHE\@.

The strength and the spatial distribution of the random potential thus have a strong impact on the IQHE and other phenomena that take place in a 2DEG\@. Unfortunately, however, the detail of the profile of the random potential is generally not well known in a GaAs/AlGaAs 2DEG\@. This is mainly because of the lack of experimental methods to probe the potential landscape in general in a 2DEG buried at the depth typically $\sim$100 nm below the surface. A notable exception in this respect arises when the potential profile possesses a unidirectional periodicity; for a unidirectional periodic potential modulation, one can map out the potential profile by analyzing the commensurability oscillation (CO) --- the magnetoresistance oscillation originating from the commensurability between the modulation period and the cyclotron radius $R_\mathrm{c}$ \cite{Weiss89,Gerhardts89,Winkler89}. The magnitude and the period of the modulation are revealed, with the aid of the first-order perturbation theory \cite{Peeters92}, from the amplitude and the period in $1/B$ of the CO, respectively. The present authors have shown \cite{Endo00e} that quantitatively reliable values of the modulation amplitude can be obtained by slightly modifying the theory to account for the scattering of the electrons out of the cyclotron orbit, which alters the damping of the CO with decreasing magnetic field \cite{Mirlin98,Endo00e}. Furthermore, by employing detailed Fourier analyses, the potential profile is shown to be accurately reconstructed up to the fourth harmonics \cite{Endo00e,Endo05HH,Endo08FCO,Endo08ModSdH}; the analysis of the CO plays analogous roles to the diffraction techniques used to determine the crystal structure.

In the present paper, we measure activation energies $\Delta /2$ of IQH states in a unidirectional lateral superlattice (ULSL) --- a 2DEG subjected to a unidirectional periodic modulation. Here the potential modulation serves as an ``artificial disorder'', or an extra source of Landau-level broadening, whose properties are well characterized through the analysis of CO\@. By comparing the activation energies with those measured in the unpatterned region of the same 2DEG wafer, we can examine the effect of the introduced modulation on the IQH states. We find that the activation energy is reduced in the ULSL by $W_N(B)$, the increment of the Landau-level half width caused by the introduction of the modulation. As will be shown below, $W_N(B)$ depends on the magnetic field $B$ owing to the commensurability effect. The decrement of $\Delta /2$ is observed to follow the magnetic-field dependence expected for $W_N(B)$.

The effect of a unidirectional periodic modulation on IQH states has been studied by several authors \cite{Muller95,Sfaxi96,Tornow96,Petit97,Milton00,Iye02JPCS,Iye02PE,Vyborny02,Feil07}. However, they are mainly interested in the strong modulation having the amplitude comparable to the cyclotron energy $\hbar \omega_\mathrm{c}$ and/or the Fermi energy $E_\mathrm{F}$. In such cases, perturbative treatment of modulation potential, which forms the basis of the quantitative analysis of CO, is not applicable and therefore the quantitative detail of the potential profile is difficult to obtain. 
In the present paper, the amplitude of the modulation amplitude is kept less than 5 percent of the $E_\mathrm{F}$, allowing the perturbation theory to describe the system accurately. 

\section{Experimental Details\label{Experimental}}
The ULSL sample used in the present study was fabricated from a GaAs/Al$_{0.3}$Ga$_{0.7}$As single-heterostructure 2DEG wafer with the heterointerface located at the depth 90 nm from the surface. The electron density $n_e$ and the mobility $\mu$ of the wafer were 2.1$\times$10$^{15}$ m$^{-2}$ and 77 m$^2$/(Vs) in the dark and increased up to 2.9$\times$10$^{15}$ m$^{-2}$ and 106 m$^2$/(Vs) after illumination by an infrared light emitting diode. The unidirectional potential modulation with a period $a=184$ nm was introduced by placing a grating of high-resolution negative electron-beam resist on the surface \cite{Endo00e}, which generates the potential modulation in the 2DEG plane via the strain-induced piezoelectric effect \cite{Skuras97}. A Hall-bar pattern (width 37 $\mu$m) with two sets of voltage probes, as depicted in the inset of Fig.\ \ref{d5015GFLF}(a), was employed, which enables simultaneous measurement of the section with (ULSL) and without (plain 2DEG) the modulation. We assume that the activation energy measured in the plain 2DEG section also represents the hypothetical activation energy in the ULSL section in the absence of the modulation, and take the difference of $\Delta /2$ between the two sections as the effect resulting exclusively from the modulation. This is not obvious due to the inevitable macroscopic inhomogeneity over the 2DEG wafers, although the two regions are only roughly one hundred $\mu$m apart. Furthermore, the microscopic spatial distributions of the random potential in the two sections are certainly not exactly the same. We have verified by the measurement of the Hall and the Shubnikov-de Haas (SdH) effect at low magnetic fields that the parameters that reflect the spatially averaged properties of the random potential, $n_e$, $\mu$, and the quantum mobility $\mu_\mathrm{Q}$, are virtually the same within the experimental uncertainty for the two sections \cite{Endo08ModSdH}, and further assumed here that the activation energy is similarly unaffected by the microscopic difference in the random potential.

The measurement was carried out in a dilution refrigerator for temperatures ranging from the base temperature ($\sim$15 mK) up to $\sim$500 mK\@. In this temperature range, the resistivity at even-integer QH states (filling factors $\nu = n_e h/(eB) = 4, 6, 8, 10 , 12$ in Figs.\ \ref{d5015GFLF} and \ref{a5015GFLF}) remains virtually zero for both the ULSL and the plain 2DEG, preventing us from determining the activation energies. Therefore we focus in the present paper on the odd-integer QH states ($\nu = 5, 7, 9, 11, 13$), where the intrinsic gap $E_g$ corresponds to the Zeeman gap $g^* \mu_\mathrm{B} B$ with $g^*$ the exchange-enhanced $g$-factor \cite{AndoOG74}. A standard low-frequency (13 Hz) ac lock-in technique was employed for the resistivity measurement with a low excitation current $I_\mathrm{ac}=0.5$ nA to avoid the current heating. 

As mentioned in \S \ref{Introduction}, the amplitudes of the introduced modulation can be revealed by the analysis of the CO\@. We found, by following the procedure detailed in ref.\ \citen{Endo08FCO}, the Fourier components, $v_1$= 0.31 meV, $v_2$= 0.10 meV, $v_3$= 0.07 meV, and $v_4$= 0.05 meV, for the profile of the modulation,
\begin{equation}
V(x)=\sum_{n \geq 1} {v_n } \cos (qnx+\varphi_n),
\label{potmodH}
\end{equation}
with $q=2\pi/a$. In what follows, we neglect the higher harmonics and approximate $V(x)$ by a sinusoidal modulation, 
\begin{equation}
V(x) = V_0 \cos(qx)
\label{potmod}
\end{equation}
with $V_0=v_1=0.31$ meV, noting that higher harmonics are rather small compared with the fundamental component. We will show below that the modulation given by eq.\ (\ref{potmod}) explains the behavior of the activation energy in the ULSL quite well.

\section{Landau Bands in a Unidirectional Lateral Superlattice \label{LLULSL}}
In this section, we review the effect that a sinusoidal modulation given by eq.\ (\ref{potmod}) exerts on the Landau levels \cite{Gerhardts89,Winkler89,Vasilopoulos89,Cui89,Zhang90,Peeters92,Shi02,Endo08ModSdH}. In an ideal 2DEG without disorder and in the absence of the modulation, the density of states (per unit area) under perpendicular magnetic field,
\begin{equation}
D (E)=\frac{1}{2\pi \ell^2}\sum_{N = 0}^\infty  {\sum_{\sigma} {\delta (E - E_{N\sigma }^0 )} },
\label{DOS0}
\end{equation}
are composed of a set of delta-function Landau levels located at 
\begin{equation}
E_{N\sigma}^0 = \left(N + \frac{1}{2}\right)\hbar\omega _\mathrm{c}  + \sigma g^* \mu_\mathrm{B} B,
\label{Energy}
\end{equation}
where $\omega_\mathrm{c} = eB/m^*$ denotes the cyclotron angular frequency with the effective mass $m^* = 0.067 m_e$ in GaAs, and $\sigma = \pm 1/2$ represents the spin and $\mu_\mathrm{B}$ the Bohr magneton. The effective $g$-factor $g^*$ is enhanced from the bare value ($|g|$=0.44 in GaAs) owing to the exchange interaction when the populations of up- and down-spin electrons differ \cite{AndoOG74}, namely at the magnetic fields at which the Fermi energy is located between spin-split Landau levels; $g^*$ is maximally enhanced at odd-integer fillings in a GaAs/AlGaAs 2DEG\@. The $1/(2\pi \ell^2)$-fold degeneracy, with $\ell = \sqrt{\hbar/(eB)}$ the magnetic length, of a Landau level reflects the translational symmetry of the system, owing to which the energy level does not depend on the position of the electrons. The introduction of the modulation alters the situation by letting the dependence of the energy level on the location of the guiding center $x_0 = -k_y \ell^2$ of the electron wave function $\phi_N(x-x_0)e^{i k_y y}/\sqrt{L_y}$, where $\phi_N(x)$ denotes the harmonic-oscillator wave function and $L_x$ (to appear below) and $L_y$ are the length of the 2DEG in the $x$ and $y$ directions, respectively. The degeneracy is thus lifted, and the density of states are now modified to
\begin{equation}
{D_W} (E) = \frac{1}{L_x L_y} \sum_{N = 0}^\infty  {\sum_{\sigma} \frac{L_y}{2\pi} \int dk_y {\delta (E - E_{N\sigma }(x_0) )} }.
\label{DOSV0}
\end{equation}
The energy levels, within the first-order perturbation theory appropriate for small $V_0/E_\mathrm{F}$, are given by
\begin{equation}
E_{N\sigma} (x_0) = E_{N\sigma}^0 + W_N(B) \cos(qx_0),
\label{EnergyV}
\end{equation}
where
\begin{equation}
W_N(B) = V_0 e^{-u/2}L_N(u),
\label{WN}
\end{equation}
$u=q^2 \ell^2/2=\pi h/(a^2 eB)$, and $L_N(u)$ is the Laguerre polynomial. The last term in eq.\ (\ref{EnergyV}) represents the change in the energy levels from those in the absence of the modulation, which varies from $-|W_N(B)|$ to $+|W_N(B)|$ depending on the position of the guiding center $x_0$ (note that $W_N(B)$ can be negative); a Landau level acquires a width $2 |W_N(B)|$ to form a ``Landau band'' through the introduction of the modulation. For transport properties, DOS at the Fermi energy plays crucial roles. We let the symbol $N_\mathrm{F} ( = [\nu/2])$ denote the index of the Landau level at which the Fermi energy is located. The width $W_{N_\mathrm{F}}(B)$ oscillates with the magnetic field due to the commensurability effect. The oscillation is the origin of the CO in the low magnetic field regime \cite{Gerhardts89,Winkler89,Vasilopoulos89,Cui89,Zhang90,Peeters92}. The commensurability effect can be made more apparent by looking at the asymptotic expression for $W_{N_\mathrm{F}}(B)$ valid at $N_\mathrm{F} \gg 1$, namely at low magnetic fields, 
\begin{equation}
\overline{W}(B) = 
V_0 \sqrt {\frac{2}{\pi q R_\mathrm{c}}} \cos \left(q R_\mathrm{c} - \frac{\pi}{4} \right),
\label{Wapp}
\end{equation}
where $R_\mathrm{c}=\hbar k_\mathrm{F}/(eB)$ is the cyclotron radius with $k_\mathrm{F}=\sqrt{2 \pi n_e}$ the Fermi wave number. Notably, the width $\overline{W}(B)$ vanishes at the flat band conditions,
\begin{equation}
\frac{2 R_\mathrm{c}}{a}=n-\frac{1}{4}\hspace{10mm}(n=1, 2, 3,...).
\label{flatband}
\end{equation}
It turns out that $\overline{W}(B)$ remains to be a good approximation of $W_{N_\mathrm{F}}(B)$ throughout the magnetic field range examined in the present paper (see Fig.\ \ref{actvda}). Therefore, $W_{N_\mathrm{F}}(B)$ is also close to zero at the flat band conditions eq.\ (\ref{flatband}).

In a real 2DEG sample, a Landau level inevitably possesses a finite width deriving from disorder. The disorder broadening of the Landau level can be conveniently modeled by replacing the delta function in eq.\ (\ref{DOS0}) by a Lorentzian \cite{Brezin84,Benedict86,Ashoori92,Potts96,Zhu03,Dial07,Shi02,Endo08SdHH,Endo09FR},
\begin{equation}
P(E) = \frac{\Gamma}{\pi}\frac{1}{E^2+\Gamma^2},
\label{LorentzP}
\end{equation}
having a width $\Gamma$ independent of the magnetic field \cite{Ashoori92,Potts96,Shi02,Endo08SdHH,Endo09FR}. Here the Lorentzian lineshape is chosen simply because it allows us to assess the effect of the modulation with analytic formulas to be shown below [eqs.\ (\ref{DOSV})-(\ref{Theta})]. The detail of the lineshape does not affect the main conclusion of the present paper. The resultant DOS for a plain 2DEG,
\begin{equation}
D(E)=\frac{1}{2\pi \ell^2}\sum_{N = 0}^\infty  {\sum_{\sigma} {P (E - E_{N\sigma }^0 )} },
\label{DOS}
\end{equation}
is plotted in Fig.\ \ref{LLs} (a). Similarly, in the presence of the modulation, DOS can be acquired by substituting the Lorentzian for the delta function in eq.\ (\ref{DOSV0}). Noting that the function to be integrated in eq.\ (\ref{DOSV0}) is the periodic function of $x_0$ with the period $a$, we can replace the integration $(1/L_x)\int dk_y = (1/\ell ^2)(1/L_x)\int_0^{L_x} dx_0$ by $(1/\ell ^2)(1/a)\int_0^a dx_0$ to obtain \cite{Endo08ModSdH}
\begin{equation}
D_W(E;W_N)=\frac{1}{2\pi \ell^2}\sum_{N = 0}^\infty  {\sum_{\sigma} {P_W (E - E_{N\sigma }^0 ;W_N)} },
\label{DOSV}
\end{equation}
where
\begin{eqnarray}
P_W(E;W_N)=\frac{1}{\pi} \int_0^\pi P(E - W_N \cos \vartheta)d\vartheta \nonumber \\
=\sqrt{\frac{1}{\pi \Gamma}} [P(E-W_N) P(E+W_N)]^{1/4} \sin\left(\frac{\Theta_+ + \Theta_- }{2}\right)
\label{VLorP}
\end{eqnarray}
with
\begin{equation}
\Theta_{\pm} = \arccos \left( \frac{E \pm W_N}{\sqrt{(E \pm W_N)^2+\Gamma^2}} \right).
\label{Theta}
\end{equation}
The peak function $P_W(E;W_N)$ is simply the superposition of the Lorentzian $P(E)$ located within the interval $[E-W_N,E+W_N]$, with heavier weight assigned toward the Landau-band edges, $\vartheta=q x_0=0$ and $\pi$. Therefore the peaks gain width $|W_N|$ in addition to the original disorder broadening $\Gamma$, and accordingly the peak width increases with $|W_N|$. At large enough $|W_N|$ ($|W_N| > \sqrt{2}\Gamma$, see Appendix), $P_W(E;W_N)$ splits into two peaks, signaling the nascent manifestation of the van Hove singularity at the band edges. On the other hand, in the limit $W_N \rightarrow 0$ (the flat band conditions), $P_W(E;W_N) \rightarrow P(E)$, and therefore $D_W(E;W_N)$ becomes identical to the DOS without modulation, $D(E)$. The DOS $D_W(E;W_N)$ for a typical value of $|W_N|$ ($> \sqrt{2}\Gamma$) is plotted in Fig.\ \ref{LLs} (b). The activation gap $\Delta /2$ is expected to be reduced by $|W_N|$ in the ULSL from the value in the plain 2DEG (irrespective of whether or not $P_W(E;W_N)$ is split), and the decrement vanishes at the flat band conditions.  

\section{Experimental Results}
\begin{figure}
\includegraphics[bbllx=20,bblly=70,bburx=575,bbury=800,width=8.5cm]{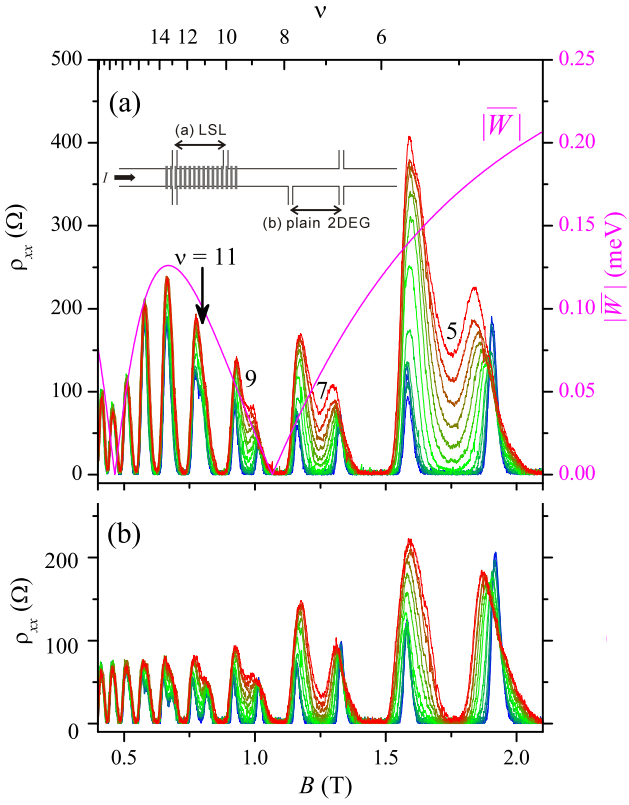}%
\caption{(Color online) Magneto-resistivity traces for (a) ULSL and (b) adjacent plain 2DEG, for temperatures $T$ (mK) = 15 (bottom, blue), 25, 35, 50, 70, 90, 120, 170, 220, 275, 325, 375, 435, and 480 (top, red), measured in dark, $n_e$=2.1$\times$10$^{15}$ m$^{-2}$. Approximated Landau bandwidth $|\overline{W}|$ calculated by eq.\ (\ref{Wapp}) is also plotted in (a) (right axis). The inset depicts the schematic of the sample. \label{d5015GFLF}}
\end{figure}

\begin{figure}
\includegraphics[bbllx=20,bblly=70,bburx=575,bbury=800,width=8.5cm]{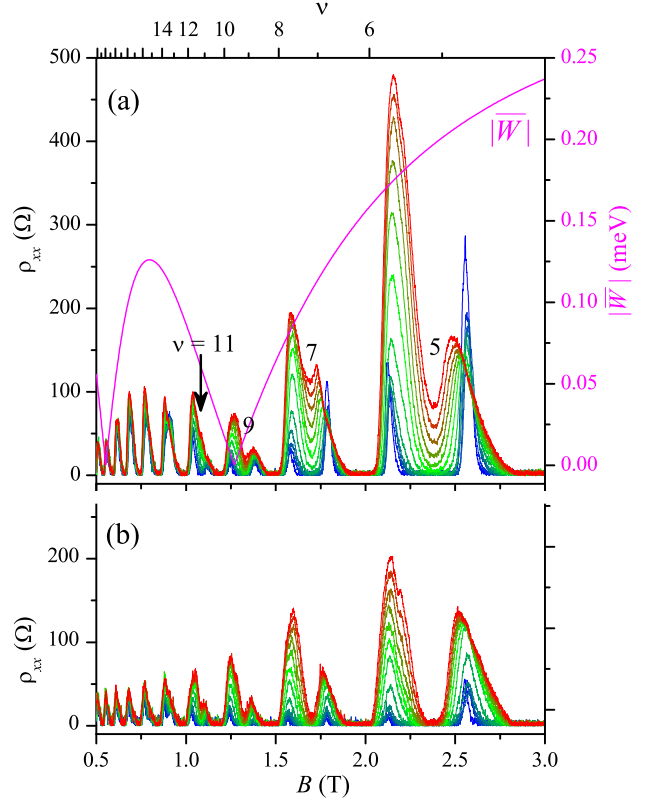}%
\caption{(Color online) Magneto-resistivity traces for (a) ULSL and (b) adjacent plain 2DEG, for temperatures $T$ (mK) = 15 (bottom, blue), 50, 70, 90, 120, 170, 220, 270, 320, 380, 420, and 490 (top, red), measured after illumination, $n_e$=2.9$\times$10$^{15}$ m$^{-2}$. Approximated Landau bandwidth $|\overline{W}|$ calculated by eq.\ (\ref{Wapp}) is also plotted in (a) (right axis).\label{a5015GFLF}}
\end{figure}

In Figs.\ \ref{d5015GFLF} and \ref{a5015GFLF}, we plot magneto-resistivity traces taken in dark ($n_e$=2.1$\times$10$^{15}$ m$^{-2}$) and after illumination ($n_e$=2.9$\times$10$^{15}$ m$^{-2}$), respectively, at different temperatures as noted in the figure captions. In both figures, the top (a) and the bottom (b) panels represent the ULSL and the adjacent plain 2DEG, respectively. In the top panels, we also plot the (approximate) Landau bandwidth $|\overline{W}|$ calculated using eq.\ (\ref{Wapp}) as an eye-guide. It can readily be perceived that the heights of the SdH peaks (or inter-quantum-Hall transition peaks) for the ULSL follow the trend of the Landau bandwidth; the peak heights are almost the same as those for the plain 2DEG at the flat band conditions (, where the effect of the modulation virtually vanishes), while the peaks grow much higher at the magnetic field where the bandwidth is larger. The latter is attributable to the diffusion (or band) contribution of the modulation potential to the SdH effect \cite{Peeters92}, as described in our previous publication \cite{Endo08ModSdH}.

\begin{figure}
\includegraphics[bbllx=0,bblly=10,bburx=550,bbury=380,width=8.5cm]{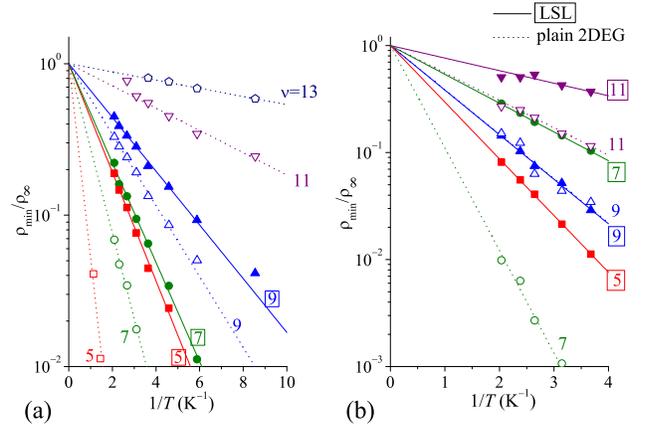}%
\caption{(Color online) Arrhenius plots of resistivity minima (normalized by $\rho_{\infty}$) at odd-integer quantum Hall states with the filling factors noted in the figure (Numbers in a square denote the filling factors for the ULSL). Solid and open symbols represent $\rho_\mathrm{min}/\rho_{\infty}$ for the ULSL and the adjacent plain 2DEG, with fits to eq.\ (\ref{rhomin}) depicted by solid and dotted lines, respectively. Only higher temperature ranges where activated transport  is observed are taken into consideration and plotted in the figures. (a) Dark, $n_e$=2.1$\times$10$^{15}$ m$^{-2}$. (b) After illumination, $n_e$=2.9$\times$10$^{15}$ m$^{-2}$. \label{arrda}}
\end{figure}

In the present paper, we focus on the odd-integer QH states. It is clear from Figs. \ref{d5015GFLF} and \ref{a5015GFLF} that the odd-integer QH states are made less robust by the introduction of the modulation; the resistivity gains a finite value at a lower temperature and increases rapidly with temperature. To be more quantitative, we examine the activation energies. The activation energy $\Delta /2$ is obtained by fitting the temperature dependence of the resistivity minima $\rho_\mathrm{min}$ to 
\begin{equation}
\rho_\mathrm{min} (T)= \rho_{\infty} \exp \left(-\frac{\Delta}{2 k_\mathrm{B} T} \right).
\label{rhomin}
\end{equation}
We present in Fig.\ \ref{arrda} the Arrhenius plot of  $\rho_\mathrm{min}$ (normalized by the extrapolated value $\rho_\infty$ in the high temperature limit) for odd-integer QH states that exhibit discernible temperature dependence in the temperature range investigated in the present study, for $n_e=2.1\times 10^{15}$ m$^{-2}$ (a) and $n_e=2.9\times 10^{15}$ m$^{-2}$ obtained after illumination (b). In the figure, we also plot the fit to eq.\ (\ref{rhomin}). Solid symbols and solid lines (open symbols and dotted lines) are for the ULSL (the plain 2DEG). It is apparent that the slope of the fitted line, which represents the activation energy, is less steep in the ULSL compared with that for the same filling factor $\nu$ in the plain 2DEG --- with a notable exception seen for $\nu=$ 9 in (b), where the slopes do not differ appreciably with and without the modulation.

\begin{figure}
\includegraphics[bbllx=20,bblly=70,bburx=500,bbury=800,width=8.5cm]{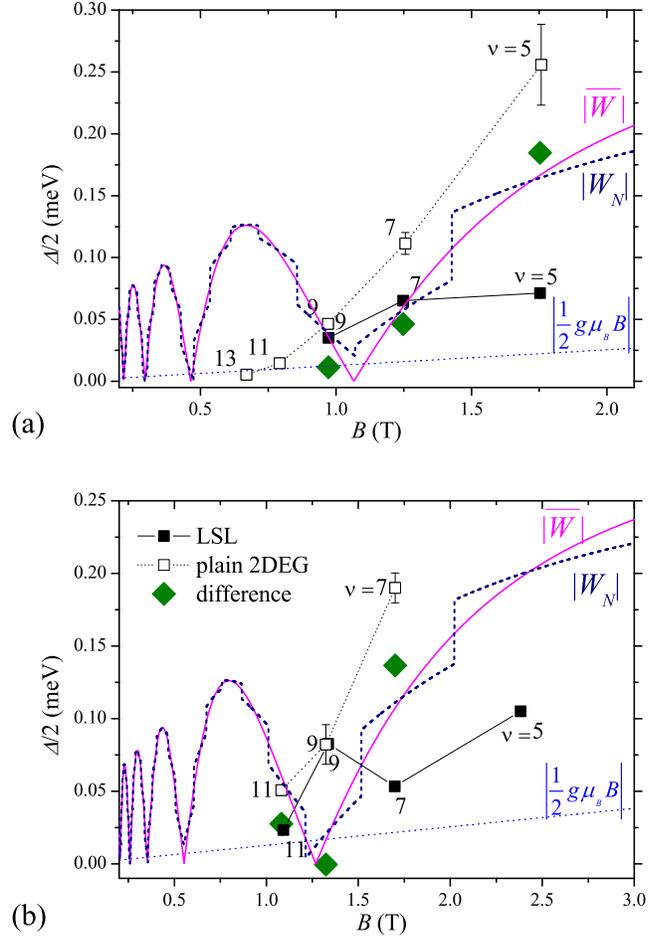}%
\caption{(Color online) Activation energy $\Delta /2$ plotted against the magnetic field, for odd-integer quantum Hall states with the filling factors noted in the figure. Solid and open squares represent the ULSL and the adjacent plain 2DEG, respectively. The error bars indicate the standard error in the fitting to eq.\ (\ref{rhomin}), and are smaller than the size of the symbols where the bars are not visible. The solid diamonds denote the differences in $\Delta /2$ with and without the modulation, $\Delta_\mathrm{plain 2DEG} /2-\Delta_\mathrm{ULSL} /2$.  Landau bandwidth $|W_{N_\mathrm{F}}|$ calculated by eq.\ (\ref{WN}) and that approximated by eq.\ (\ref{Wapp}), $|\overline{W}|$, are plotted by  thick dashed and solid lines, respectively. The (half) Zeeman energy $|g \mu_\mathrm{B} B / 2|$ with the bare $g$-factor ($g$=$-$0.44) is also plotted (thin dotted line). (a) Dark, $n_e$=2.1$\times$10$^{15}$ m$^{-2}$. (b) After illumination, $n_e$=2.9$\times$10$^{15}$ m$^{-2}$. \label{actvda}}
\end{figure}

The activation energy thus obtained is plotted in Fig.\ \ref{actvda} against the magnetic field at which the QH state takes place. Again (a) and (b) are for $n_e=2.1\times 10^{15}$ m$^{-2}$ and $n_e=2.9\times 10^{15}$ m$^{-2}$, respectively, and solid and open squares represent the ULSL and the plain 2DEG, respectively. Owing to the exchange enhancement, the activation energies $\Delta /2$ are larger than (the half of) the bare Zeeman energy $|g \mu_\mathrm{B} B/2|$, even in the presence of the modulation. The width of the Landau band $|W_{N_\mathrm{F}}|$, eq.\ (\ref{WN}), is also plotted, along with its asymptotic expression $|\overline{W}|$, eq.\ (\ref{Wapp}). The effect of the modulation on the activation energy is extracted by subtracting $\Delta /2$ of the ULSL from that of the plain 2DEG\@. The decrement in the activation energy due to the modulation $\Delta_\mathrm{plain2DEG}/2-\Delta_\mathrm{ULSL}/2$, plotted by solid diamonds in the figures, is seen to coincide well with the Landau band width; the activation energy is reduced by an amount equal to the increment of the Landau band width brought about by the introduction of the modulation. This is exactly what we expected in the previous section, \S \ref{LLULSL}. The filling factor $\nu=$ 9 in (b) happens to fall in the vicinity of a flat band condition $|W_N|\sim 0$, and therefore the activation energy there does not change noticeably by the introduction of the modulation.

\section{Discussion\label{Discussion}}
\begin{figure}
\includegraphics[bbllx=0,bblly=0,bburx=560,bbury=610,width=8.5cm]{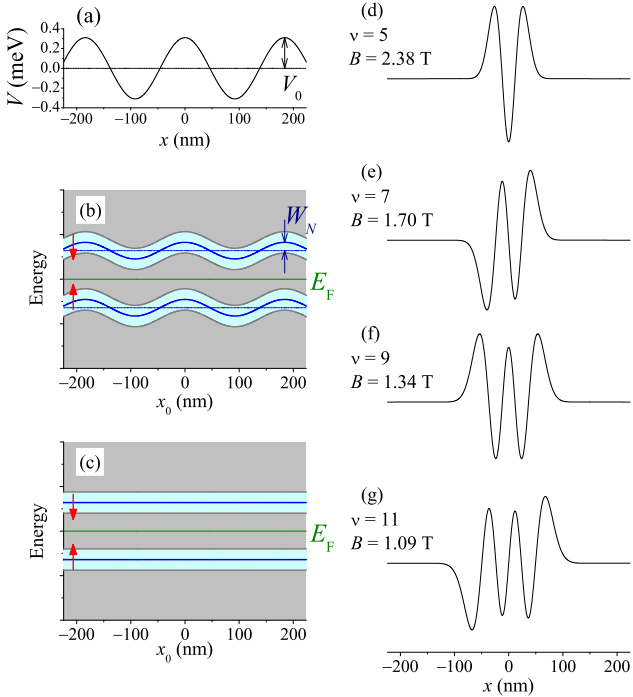}%
\caption{(Color online) (a) Profile of the potential modulation eq.\ (\ref{potmod}). (b)(c) Schematic drawing of the $x_0$ dependent energy levels for an odd-integer QH state. The energy levels $E_{N,\sigma}(x_0)$ given by eq.\ (\ref{EnergyV}) for both up and down spins are plotted by solid (blue) lines, sandwiched by lightly shaded (light blue) bands representing the extended states within the  disorder broadened Landau levels, for finite $W_N$ (b) and for a flat band condition $W_N=0$ (c). (d)-(g) Wave functions for $\nu=$ 5, 7, 9, 11, in the Landau gauge at the magnetic fields corresponding to $n_e=$ 2.9$\times$10$^{15}$ m$^{-2}$. The wave functions are depicted with the same horizontal scale as the modulation profile in (a). \label{perWF}}
\end{figure}
In \S \ref{LLULSL}, we evaluated the broadening of the Landau levels due to the periodic modulation, and the resulting reduction of the gap, by taking the spatial average of the energy levels (see eqs.\ (\ref{DOSV}) and (\ref{VLorP}), and the discussion just above them). As illustrated in Fig.\ \ref{perWF} (b), however, the \textit{local} gap is not altered by the modulation, since the levels both just above and below the Fermi level, possessing the same Landau index $N$ for odd-integer QH states, vary synchronously with $x_0$ in accordance with eq.\ (\ref{EnergyV}), with the same amplitude $W_N$. The averaging is justified when the spatial extent of the wave functions $\sim R_\mathrm{c} = \sqrt{2N+1} \ell$ is larger or at least comparable to the modulation period $a$. In Fig.\ \ref{perWF}, we compare the wave functions (d)-(g) with the potential modulation (a), depicted with the same horizontal scale. The wave functions are calculated at the magnetic fields corresponding to $n_e$=2.9$\times$10$^{15}$ m$^{-2}$; they slightly expand for $n_e$=2.1$\times$10$^{15}$ m$^{-2}$. It can be seen that the spatial extents of the wave functions are actually about the same order of magnitude as the modulation period. This is also obvious from the occurrence of the commensurability effect, which requires $R_\mathrm{c} \sim a$. If the QH state to be considered takes place at a much higher magnetic field and $R_\mathrm{c}$ is much smaller than $a$ (as is the case when $n_e$ is much larger), the gap felt by the electrons will be closer to the \textit{local} gap and therefore the effect of the modulation will be less prominent; the QH states are less sensitive to the modulation that varies much slower than the spatial extent of the wave functions. In fact, we have observed the $\nu$=$4/3$ fractional QH state taking place at $\sim$9 T to withstand the modulation having $W_N$ larger than the activation energy $\Delta/2$ that the state had in the plain 2DEG \cite{Endo09FQHEMOD}. Similar decrease in the sensitivity with the increase of the magnetic field was discussed for the random potential fluctuations induced by the disorder \cite{Usher90}.

In the present paper, we limited ourselves to the unidirectional LSL with the direction of the modulation parallel to the direction of the current (the $x$ direction). The direction was chosen from the necessity of using CO to deduce the modulation amplitude quantitatively. We expect that the effect of the oscillating Landau bandwidth discussed in the present paper does not depend on the direction of the modulation in the low magnetic-field regime where the bandwidth can be viewed as the spatial average. At higher magnetic fields, however, the spatial averaging is not appropriate, as discussed in the previous paragraph, and the spatial variation of the energy levels should be taken into consideration. In such magnetic field regime, the direction of the modulation bears more importance, which will be the subject of our future study. 

\section{Conclusions\label{Conclusion}}
We have examined the effect that a unidirectional periodic potential modulation exerts on the activation energy of the integer quantum Hall state. Unlike native disorders unavoidably present in the 2DEGs, the artificially introduced periodic modulation allows us to probe its strength and spatial extent through the analysis of the commensurability oscillation. Owing to the modulation, the Landau levels gain additional width, resulting in the decrease in the activation energies. The measured decrement in the activation energy due to the modulation is found to be in quantitative agreement with the increment in the Landau bandwidth calculated using the known profile of the modulation. The decrement in the activation energy (increment in the Landau-level width) varies with the magnetic field owing to the commensurability effect, and vanishes at the flat band conditions.

\section*{Acknowledgment}
This work was supported by Grant-in-Aid for Scientific Research (C) (18540312) and (A) (18204029) from the Ministry of Education, Culture, Sports, Science and Technology (MEXT).

\appendix
\section{Lineshape of the Landau Peak Function $P_W(E;W_N)$}
In this appendix, we examine in more detail the lineshape of the Landau band, given by $P_W(E;W_N)$ in eq.\ (\ref{VLorP}), in the presence of a unidirectional periodic potential modulation. For this purpose, we take the derivative of $P_W(E;W_N)$ with respect to $E$:
\begin{eqnarray}
\lefteqn{{P_W}^\prime (E;W_N) \equiv \frac{\partial P_W(E;W_N)}{\partial E} =} \nonumber \\
 & \displaystyle{-\frac{1}{2}\sqrt{\frac{\pi}{\Gamma}}[P(E-W_N)P(E+W_N)]^{1/4} \times} \nonumber \\
 & \displaystyle{\left\{ \left[ P(E-W_N)+P(E+W_N) \right] \cos\left( \frac{\Theta_+ + \Theta_-}{2} \right)\right.} \nonumber \\
 & \displaystyle{+\frac{(E-W_N)P(E-W_N)+(E+W_N)P(E+W_N)}{\Gamma}} \nonumber \\
 & \displaystyle{\left. \times\sin\left( \frac{\Theta_+ + \Theta_-}{2} \right)\right\}}.
\end{eqnarray}
It can readily be seen that ${P_W}^\prime (0;W_N)=0$, and therefore $E=0$ is a stationary point of $P_W(E;W_N)$. This is also obvious by noting that $P_W(E;W_N)$ and ${P_W}^\prime (E;W_N)$ are even and odd functions of $E$, respectively: $P_W(-E;W_N)=P_W(E;W_N)$ and ${P_W}^\prime (-E;W_N)=-{P_W}^\prime (E;W_N)$. The function $P_W(E;W_N)$ takes either local maximum or local minimum at $E=0$, depending on whether ${P_W}^{\prime \prime} (0;W_N) < 0$ or ${P_W}^{\prime \prime} (0;W_N) > 0$, where ${P_W}^{\prime \prime} (E;W_N) \equiv \partial^2 P_W(E;W_N)/\partial E^2$. Expanding ${P_W}^\prime (E;W_N)$ in the Maclaurin series, we have
\begin{eqnarray}
\lefteqn{{P_W}^\prime (E;W_N) =} \nonumber \\
 & \displaystyle{\frac{{W_N}^2-2 \Gamma^2}{\pi ({W_N}^2+\Gamma^2)^{5/2}}E+\frac{3 {W_N}^4-24 {W_N}^2 \Gamma^2+8 \Gamma^4}{2 \pi ({W_N}^2+\Gamma^2)^{9/2}} E^3} \nonumber \\
 & \displaystyle{+O(E^4)},
\end{eqnarray}
and therefore,
\begin{equation}
{P_W}^{\prime \prime} (E;W_N) = \frac{{W_N}^2-2 \Gamma^2}{\pi ({W_N}^2+\Gamma^2)^{5/2}}+O(E^2).
\end{equation}
Thus, for $W_N < \sqrt{2} \Gamma$, $E=0$ is the (local) maximum and $P_W(E;W_N)$ is comprised of a single peak, while for $W_N > \sqrt{2} \Gamma$, $P_W(E;W_N)$ splits into two peaks.

\bibliography{qhe,twodeg,lsls,ourpps,Noteiqhem,magmod,FundRel}

\begin{thebibliography}{10}
\bibitem{NoteVRH}
At very low temperatures, the conductivity is rather attributable to the
  variable-range hopping mechanism, with the temperature dependece $\sigma_{xx}
  \propto \exp [-(T_0/T)^\gamma]$. See, e.g., Refs. \citen{PrangeGirvin90} and
  \citen{Furlan98}.
\end{thebibliography}

\end{document}